\documentclass{article}
\usepackage{clpsref}
\usepackage{epsfig}
\newcommand{\wvp}{{\vec P}}
\newcommand{\vp}{{\vec p}}
\newcommand{\lsim}{\mathrel{\rlap{\lower4pt\hbox{\hskip0pt$\sim$}}\raise1pt\hbox{$<$}}}           
\newcommand{\gsim}{\mathrel{\rlap{\lower4pt\hbox{\hskip0pt$\sim$}}\raise1pt\hbox{$>$}}}           
\begin{document}
%
\title{\vspace*{-6ex}}
\author{\bf\large Pair creation and plasma oscillations
\footnote{To appear in the Proceedings of {\em Quark Matter in Astro- and
Particlephysics}, a workshop at the University of Rostock, Germany,
November 27 - 29, 2000. Eds. D. Blaschke, G. Burau, S.M. Schmidt.}\\[2ex]
\normalsize A.V. Prozorkevich and D.V. Vinnik\\[-0.5ex]
\it\small Physics Department, Saratov State University, \\[-0.5ex]
\it\small 410071 Saratov, Russian Federation\\[2ex]
\normalsize S.M. Schmidt\\[-0.5ex]
\it\small Fachbereich Physik, Universit\"at Rostock, \\[-0.5ex] \it\small
D-18051 Rostock, Germany\\[2ex]
\normalsize  M.B. Hecht and C.D. Roberts\\[-0.5ex]
\it\small Physics Division, Building 203, Argonne National Laboratory,\\[-0.5ex]
\it\small Argonne, IL 60439-4843, USA}

%
\date{\parbox{21em}{\small We describe aspects of particle creation in strong
fields using a quantum kinetic equation with a relaxation-time approximation
to the collision term.  The strong electric background field is determined by
solving Maxwell's equation in tandem with the Vlasov equation.  Plasma
oscillations appear as a result of feedback between the background field and
the field generated by the particles produced.  The plasma frequency depends
on the strength of the initial background field and the collision frequency,
and is sensitive to the necessary momentum-dependence of dressed-parton
masses}.\\[1ex]
{\it\small Pacs Numbers}\small : 05.20.Dd, 25.75.Dw, 05.60.Gg,
12.38.Mh\vspace*{-4ex}}
\maketitle

%

%
Ultra-relativistic heavy-ion collisions are complicated processes and their
understanding requires a microscopic modelling of all stages: the formation,
evolution and hadronisation of a strongly coupled plasma. If the energy
density produced in the interaction volume is large enough, then the relevant
degrees of freedom are quarks and gluons. The terrestrial recreation of this
quark gluon plasma (QGP) will aid in understanding phenomena such as the big
bang and compact stars.

Construction of the Relativistic Heavy Ion Collider at the Brookhaven
National Laboratory is complete and the initial energy density: $\varepsilon
\sim 10-100$ GeV/fm$^3$, expected to be produced in the collisions at this
facility is certainly sufficient for QGP formation.  Experimentally there are
two parameters that control the conditions produced: the beam/target
properties and the impact parameter.  Varying these parameters changes the
nature of the debris measured in the detectors.  Signals of QGP formation and
information about its detailed properties are buried in that
debris~\cite{signals}.  Predicting the signals and properties requires a
microscopic understanding of the collisions, including their non-equilibrium
aspects.

In the space-time evolution of a relativistic heavy ion collision the initial
state is a system far from equilibrium.  This system then evolves to form an
equilibrated QGP, and the investigation of that evolution and the signals
that characterise the process are an important contemporary aspect of QGP
research.

The formation of a QGP is commonly described by two distinct mechanisms: the
perturbative parton picture~\cite{Gribovpartons} and the string
picture~\cite{anders83}.  In the parton picture the colliding nuclei are
visualised as clouds of partons and the plasma properties are generated by
rapid, multiple, short-range parton-parton interactions.  In the string
picture the nuclei are imagined to pass through one another and stretch a
flux tube between them as they separate, which decays via a nonperturbative
particle-antiparticle production process.  These approaches are complimentary
and both have merits and limitations.  Once the particles are produced the
subsequent analysis proceeds using Monte-Carlo event generators
\cite{partonMC,Geiger95,stringMC}.

\begin{figure}[t]
\centering{\ \epsfig{figure=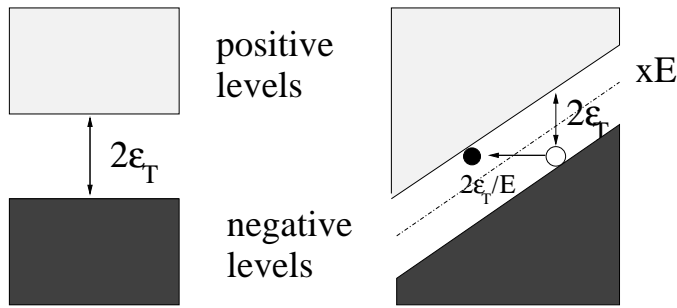,height=4cm}}\vspace*{2ex}

\caption{\label{fig1} {\em Left}: For $E=0$ the vacuum is characterised by a
completely filled negative-energy Dirac sea and an unoccupied positive-energy
continuum, separated by a gap: $2\, \varepsilon_{\rm T}= 2\,(m^2 +
p_\perp^2)^{1/2}$. {\em Right}: Introducing a constant external field:
$\vec{E} = \hat{e}_x E$, which is produced by a potential: $A^0= -E \vec{x}$,
tilts the energy levels.  In this case a particle in the negative-energy sea
will tunnel through the gap with a probability $\sim
\exp(-\pi\varepsilon_T^2/eE)$.  Succeeding, it will be accelerated by the
field in the $-x$-direction, while the hole it leaves behind will be
accelerated in the opposite direction.  The energy-level distortion is
increased with increasing $E$ and hence so is the tunneling probability.
$eE$ can be related to the flux-tube string-tension.}
\end{figure}

Herein we employ the nonperturbative flux tube picture \cite{nuss}.  A flux
tube is characterised by a linearly rising, confining quark-antiquark
potential: $V_{q{\bar q}}(r)=\sigma\,r$.  The string tension can be estimated
in lattice simulations using static quark sources, which yields $\sigma \sim
4 \Lambda_{\rm QCD}^2\sim 1\,$GeV$/$fm.  This string tension can be viewed as
a strong background field that destabilises the vacuum and the instability is
corrected through particle-antiparticle production via a process akin to the
Schwinger mechanism~\cite{Sch}.  Figure~\ref{fig1} is an artist's impression
of this process.

Assuming a constant, uniform, Abelian field, $E$, one is able to derive an
expression for the rate of particle production via this nonperturbative
mechanism
\begin{equation}
S(p_\perp)=\frac{dN}{dtdVd^2p_\perp}=|eE|\ln
\bigg[1+\exp\bigg(-\frac{2\pi(m^2+p_\perp^2)}{|eE|}\bigg)\bigg].
\end{equation}
where $m$ is the mass and $e$ the charge of the particles produced.  It is
plain from this equation that the production rate is enhanced with increasing
electric field and suppressed for a large mass and/or transverse momentum.

The production of charged particles leads naturally to an internal
current.  
That current produces an electric field
that increasingly screens and finally completely neutralises the background
field so that particle creation stops.  However, the current persists and the
field associated with that internal current restarts the production process
but now the field produced acts to retard and finally eliminate the
current. \ldots This is the back-reaction phenomenon and the natural
consequences are time dependent fields and currents~\cite{Back}.  One
observable and necessary consequence is plasma oscillations.  Their
properties, such as frequency and amplitude, depend on the initial strength
of the background field and the frequency of interactions between the
partons.  It is clear that very frequent collisions will rapidly damp plasma
oscillations and equilibrate the system.

The process we have described can be characterised by four distinct
time-scales:\\[-4.0ex]
\begin{enumerate}
\item the quantum time, $\tau_{qu}$, which is set by the Compton wavelength
of the particles produced and identifies the time-domain over which they can
be localised/identified as ``particles;''\vspace*{-1.5ex}
\item the tunneling time, $\tau_{tu}$, which is inversely proportional to the
flux tube field strength and describes the time between successive tunnelling
events;\vspace*{-1.5ex}
\item the plasma oscillation period, $\tau_{pl}$, which is also inversely
proportional to field strength in the flux tube but is affected by other
mechanisms as well; \vspace*{-1.5ex}
\item and the collision period, $\tau_{coll}$, which is the mean time between
two partonic collision events.\vspace*{-1.5ex}
\end{enumerate}
Each of the time-scales is important and the behaviour of the plasma depends
on their relation to each other~\cite{review,kme}; e.g., non-Markovian
(time-nonlocality) effects are dramatic if $\tau_{qu}\sim\tau_{tu}$.  Herein
we focus on the interplay of the larger time scales and consider a system for
which $\tau_{pl}\sim\tau_{coll}$.

Particle creation is a natural outcome when solving QED in the presence of a
strong external field and a formulation of this problem in terms of a kinetic
equation is useful since, e.g., transport properties are easy to explore.
The precise connection between this quantum kinetic equation and the mean
field approximation in non-equilibrium quantum field theory \cite{neq} is not
trivial and the derivation yields a kinetic equation that, importantly, is
non-Markovian in character~\cite{Rau,basti,gsi,prd}.

The single particle distribution function is defined as the vacuum
expectation value, in a time-dependent basis, of creation and annihilation
operators for single particle states at time $t$ with three-momentum
$\vec{p}$: $a^\dagger_{\vec{p}}(t)$, $a_{\vec{p}}(t)$; i.e., 
\begin{equation}
f(\vec{p},t):= \langle 0 | a^\dagger_{\vec{p}}(t)\,a_{\vec{p}}(t)|0\rangle\,.
\end{equation}
The evolution of this distribution function is described by the following
quantum Vlasov equation:
\begin{eqnarray}
\label{10}
\lefteqn{
\frac{df_\pm(\wvp,t)}{dt}=\frac{\partial f_\pm(\wvp,t) }{\partial
t}+eE(t)\frac{\partial f_\pm(\wvp,t)}{\partial P_\parallel(t)} }&&\\
\nonumber &=&\frac{1}{2}{\cal W}_\pm(t)\int_{-\infty}^t dt'{\cal
W}_\pm(t')\times [1 \pm 2 f_\pm(\wvp,t')]\cos[x(t',t)]+C_\pm(\wvp,t)\,,
\end{eqnarray}
where the lower [upper] sign corresponds to fermion [boson] pair creation,
$C$ is a collision term and ${\cal W}_\pm$ are the transition amplitudes.
The momentum is defined as $\wvp=(p_1,p_2,P_\parallel(t))$, with the
longitudinal [kinetic] momentum $P_\parallel(t)=p_3-eA(t)$.

We approximate the collision-induced background field by an external,
time-dependent, spatially homogeneous vector potential: $A_\mu$, in Coulomb
gauge: $A_0=0$, taken to define the $z$-axis: ${\vec A} = (0,0,A(t))$.  The
corresponding electric field, also along the $z$-axis, is
\begin{equation}
E(t) = -{\dot A}(t)=-\frac{dA(t)}{dt}\,.
\end{equation}

For fermions~\cite{Rau,gsi} and bosons~\cite{kme,gsi} the transition
amplitudes are
\begin{equation}
\label{12}
{\cal W}_-(t)=\frac{eE(t)\varepsilon_\perp}{\omega^2(t)}\,,\;\;
{\cal W}_+(t)=\frac{eE(t)P_\parallel(t)}{\omega^2(t)}\,,
\end{equation}
where the transverse energy
$\varepsilon_\perp=\sqrt{m^2+\vec{p}_\perp^{\,2}}$, $\vp_\perp = (p_1,p_2)$,
and $\omega(t)=\sqrt{\varepsilon_\perp^2+P_\parallel^2(t)}$ is the total
energy.  In Eq.~(\ref{10}),
\begin{equation}
\label{30} 
x(t^\prime,t) = 2[\Theta(t)-\Theta(t^\prime)]\,,\;\;
\Theta(t) = \int^t_{-\infty}dt^\prime \omega(t^\prime)\,,
\end{equation}
is the dynamical phase difference.

The time dependence of the electric field is obtained by solving the Maxwell
equation
\begin{equation}
\label{Edot}
-{\ddot A}^\pm(t)= {\dot E}^\pm(t)= - j^{ex}(t) - j_{cond}(t) - j_{pol}(t)
\end{equation}
where the three components of the current are: the external current
generated, obviously, by the external field; the conduction current
\begin{equation}
\label{jcond}
j_{cond}(t)= g_\pm
e\int\!\frac{d^3p}{(2\pi)^3}\,\frac{P_\parallel(t)}{\omega(\wvp,t)}
f_\pm(\wvp,t) \,,
\end{equation}
associated with the collective motion of the charged particles; and the
polarisation current
\begin{equation}
\label{jpol}
j_{pol}(t) = g_\pm e \int \!\!  \frac{d^3P}{(2\pi)^3}
\frac{P_\parallel(t)}{\omega(\wvp,t)} \left[ \frac{S(\wvp,t)}{{\cal
W}_\pm(\wvp,t)}-\frac{e\, \dot E^\pm(t)\,P_\parallel(t) }
{8\,\omega^4(\vec{P},t)} \right]
\bigg(\frac{\epsilon_\perp}{P_\parallel(t)}\bigg)^{g_\pm-1}\!\!\!,
\end{equation}
which is proportional to the production rate.  Here $g_-=2$, $g_+=1$ and all
fields and charges are understood to be fully renormalised, which has a
particular impact on the polarisation current~\cite{bloch}.

\begin{figure}[t]
\centerline{\epsfig{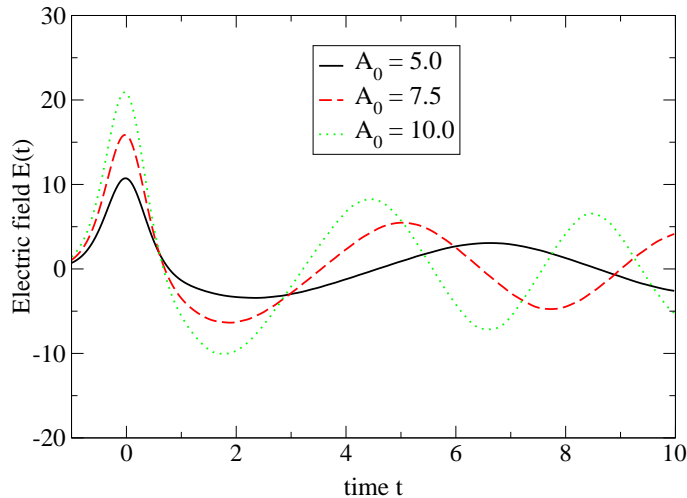}}
\caption{Time evolution of the fermion's electric field obtained with three
different initial field strengths.  The amplitude and frequency of the plasma
oscillations increases with an increase in the strength of the external
impulse field, Eq.~(\protect\ref{Eext}).  The reference time-scale is the
lifetime $b$ of the impulse current.  (The field and time are given in
arbitrary units, and $b=0.5$.)\label{fig2}}
\end{figure}

We mimic a relativistic heavy ion collision by using an impulse profile for
the external field
\begin{equation}
\label{Eext}
E_{ex}=-\frac{A_0}{b} {\rm sech}^2(t/b)\,,
\end{equation}
which is our two-parameter model input: the width $b$ and the amplitude $A_0$
are chosen so that the initial conditions are comparable to
typical/anticipated experimental values.  This is the seed in a solution of
the coupled system of Eqs.~(\ref{10}) and (\ref{Edot})--(\ref{jpol}), and in
Fig.~\ref{fig2} we illustrate the result for the electric field {\it in the
absence of collisions}.

The qualitative features are obvious and easy to understand.  The external
impulse (the collision) is evident: the portion of the curve roughly
symmetric about $t=0$, where the field assumes its maximum value $A_0/b$.  It
produces charged particles and accelerates them, producing a positive current
and an associated field that continues to oppose the external field until the
net field vanishes.  At that time particle production ceases and the current
reaches a maximum value.  However, the external field is dying away: it's
lifetime is $t\sim 0.5$, so that the part of the electric field due to the
particles' own motion quickly finds itself too strong.  The excess of field
strength begins to produce particles.  It accelerates these in the opposite
direction to the particles generating the existing current whilst
simultaneously decelerating the particles in that current.  That continues
until the particle current vanishes, at which point the net field has
acquired its largest negative value.  Particle production continues and with
that a negative net current appears and grows. \ldots\ Now a pattern akin to
that of an undamped harmonic oscillator has appeared.  In the absence of
other effects, such as collisions, it continues in a steady state with the
magnitude and period of the plasma oscillations determined by the two model
parameters that characterise the collision.

The quantum kinetic equation, Eq.~(\ref{10}), describes a system far from
equilibrium and therefore any realistic collision term $C$ valid shortly
after the impact must be expected to have a very complex form.  However, as
Fig.~\ref{fig2} illustrates, the evolution at not-so-much later times is
determined by the properties of the particles produced and not by the violent
nonequilibrium effects of the collision.  This observation suggests that the
parton plasma can be treated as a quasi-equilibrium system and that the
effects of collisions can be represented via a relaxation time
approximation~\cite{bloch,memory,eis,hydro,nayak2}:
\begin{equation}
C(\vp,t)=\frac{1}{\tau(t)}[f_-^{eq}(\vp,t)-f(\vp,t)]\,,
\end{equation}
with $\tau(t)$ an in general time-dependent ``relaxation time'' (although we
use a constant value $\tau(t)=\tau_r$ in the calculations reported herein)
where
\begin{equation}
\label{eq}
f_-^{eq}(\vp,T(t))= \left[ \exp\left( \frac{p_{\nu}u^{\nu}(t)}{T(t)}\right) + 1
\right]^{-1}
\end{equation}
is the quasi-equilibrium distribution function for fermions.  Here $T(t)$ is
a local-temperature and $u^\nu(t)$, $u^2=1$, is a hydrodynamical
velocity~\cite{hydro}.  (With our geometry, $u^{\nu}(t) =
(1,0,0,v(t))/[1-v^2(t)]^{1/2}$.)

Our definition of quasi-equilibrium is to require that at each $t$ the energy
and momentum density in the evolving plasma are the same as those in an
equilibrated plasma; i.e., we require that
\begin{equation}
\label{eneneq}
\epsilon_f(t) = \epsilon^{eq}(t)\,,\; \vec{p}_f(t) = \vec{p}^{\,eq}(t)
\end{equation}
where, as one would expect, 
\begin{equation}
\epsilon^{eq}(t)= \int\frac{d^3p}{(2\pi)^3}\,\omega(\vp,t)\,
f_-^{eq}(\vp,t)\,,\; 
\vec{p}^{\,eq}(t)= \int\frac{d^3p}{(2\pi)^3}\, \vec{p}(t)\,f_-^{eq}(\vp,t)\,,
\end{equation}
and 
\begin{eqnarray}
\epsilon_f(t) & = & \int\frac{d^3p}{(2\pi)^3}\,\omega(\vp,t)\left[f_-(\vp,t)-
z_{f_-}(\vp,t)\right] \,, \\
\vec{p}_f(t) & = &\int\frac{d^3p}{(2\pi)^3}\,\vec{p}(t)\left[f_-(\vp,t)-
z_{f_-}(\vp,t)\right]\,,
\end{eqnarray}
where 
\begin{equation}
z_{f_-}(\vp,t) = \bigg(\frac{e\varepsilon_{\perp}}{4 \omega^3}\bigg)^2\bigg[
E(t)^2 - \frac{e^{-2t/\tau_r}}{\tau_r}\int \limits_{-\infty}^{t}dt' E(t')^2
e^{2t'/\tau_r} \bigg]
\end{equation}
is a regularising counterterm.  Adding Eqs.~(\ref{eneneq}) to the system of
coupled equations embeds implicit equations for the temperature profile,
$T(t)$, and collective velocity, $v(t)$.\footnote{A shortcoming of the
approach presented thus far is that it neglects the possibility of
dissipative inelastic scattering events transforming electric field energy
directly into temperature.  However, we have almost completed an extension
valid in that case.}

Solving the complete set of coupled equations is a straightforward but time
consuming exercise.  For the present illustration we employ the minor
simplification of assuming that the equilibrium energy density is that of a
two-flavour, massless, free-quark gas; i.e.,
\begin{equation}
\epsilon^{eq}(T(t)) = \frac{7\pi^2}{10} T^4(t)
\end{equation}
and then proceed.  

\begin{figure}[t]
\centerline{\epsfig{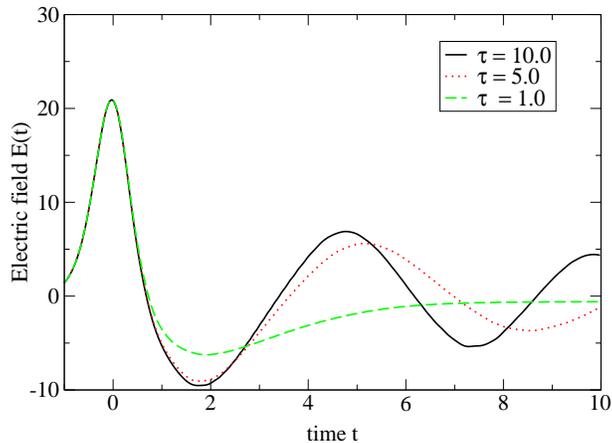}}
\caption{Time evolution of the electric field for three different relaxation
times.  As one would anticipate, collisions damp the plasma oscillations:
their frequency and amplitude decreases with decreasing relaxation time.
Solution obtained using the impulse profile, Eq.~(\protect\ref{Eext}), with
$A_0=10$, $b=0.5$.  The field and the time are given in arbitrary
units.\label{fig3} }
\end{figure}

The solution obtained for the electric field using the impulse profile,
Eq.~(\ref{Eext}), is depicted Fig.~\ref{fig3}.  Here the behaviour is
analogous to that of a damped oscillator: the frequency and amplitude of the
plasma oscillations diminishes with increasing collision frequency,
$1/\tau_r$ (friction).

In the examples presented hitherto we have employed a constant fermion mass.
However, in QCD the dressed-quark mass is momentum-dependent~\cite{cdragw,jonivar}
and that momentum-dependence is significant when $m_0\lsim \Lambda_{\rm
QCD}$, where $m_0$ is the current-quark mass.  That can be illustrated using
a simple, instantaneous Dyson-Schwinger equation model of QCD, introduced in
Ref.~\cite{hirsch}, which yields the following pair of coupled equations for
the scalar functions in the dressed-quark propagator: $S(p)=1/[i\gamma\cdot p
\,A(\vec{p}^{\,2}) + B(\vec{p}^{\,2})]$,
\begin{eqnarray}
\label{dseB}
B(\vp,t)&=&m_0 +
\eta\frac{B(\vp,t)}{\sqrt{\vp^2\,A^2(\vp,t)+B^2(\vp,t)}}\,(1-2f_-(\vec{p},t))\,,\\
\label{dseA}
A(\vp,t)&=&\frac{2B(\vp,t)}{m_0+B(\vp,t)}\,,
\end{eqnarray}
where $\eta$ is the model's mass scale.  In this preliminary, illustrative
calculation we discard the distribution function in Eq.~(\ref{dseB}).

\begin{figure}[t]
\centerline{\epsfig{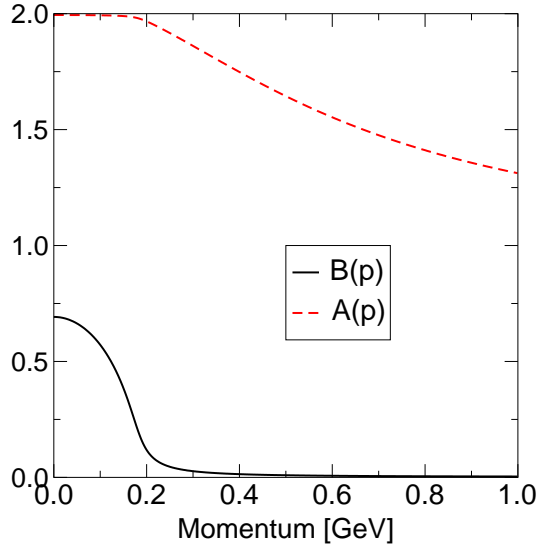}}
\caption{The scalar functions characterising a dressed-$u$-quark propagator,
as function of momentum, obtained using the simple DSE model introduced in
Ref.~\protect\cite{hirsch}, see Eqs.~(\protect\ref{dseB}),
(\protect\ref{dseA}).  We plot the functions as obtained at
$T=T_c=0.17\,$GeV, which is the critical temperature for deconfinement in the
model~\protect\cite{bastiscm}.
\label{fig4} }
\end{figure}

The solution obtained using a current-quark mass $m_0=5\,$MeV and with $\eta
= 1.33\,$GeV is depicted in Fig.~\ref{fig4}.  This value of the mass-scale
parameter can be compared with the potential energy in a QCD string at the
confinement distance, $V_{q\bar q}(r=1\,\mbox{fm}) = \sigma r \simeq (2
\Lambda_{\rm QCD})^2 (1/ \Lambda_{\rm QCD}) = 4 \Lambda_{\rm QCD}\sim
1.0\,$GeV.  The dressed-quark mass function is
$m(\vec{p}^{\,2},T)=B(\vec{p}^{\,2},T)/A(\vec{p}^{\,2},T)$, and $m(0,T)$
provides a practical estimate of the $T$-dependent constituent-quark
mass~\cite{mr97}: in this example $m(0,T_c)\approx 0.35\,$GeV.

\begin{figure}[t]
\centerline{\epsfig{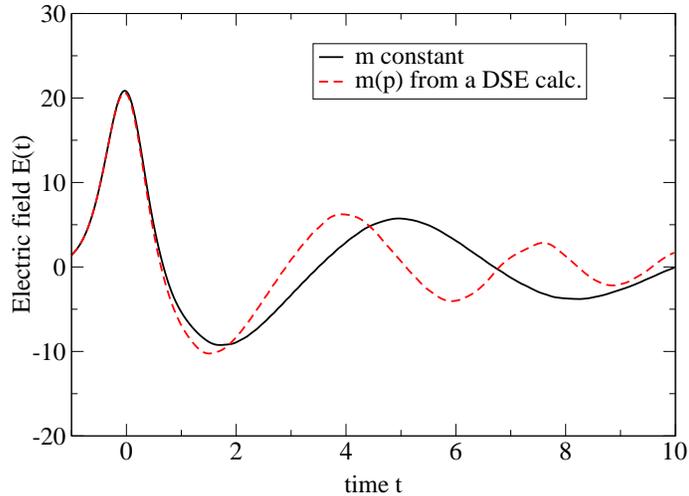}}
\caption{Electric field as function of time for $A_0=10.0$, $\tau=5.0$,
$b=0.5$.  We compare the results obtained using particles produced with a
constant mass, $m= 1$, with those obtained when the mass is
momentum-dependent.  For illustrative simplicity, we use the profile in
Fig.~\protect\ref{fig4} but normalised such that, $m(0)=m$.  The mass-scale
is arbitrary.\label{fig5} }
\end{figure}

In our flux tube model for particle production we produce fermions with
different momenta and following this discussion it is clear that the
effective mass of the particles produced must be different for each momentum.
That will affect the production and evolution of the plasma.  To illustrate
that we have repeated the last calculation using $m(p)$ wherever the particle
mass appears in the system of coupled equations for the single particle
distribution function.  The effect on the electric field is depicted in
Fig.~\ref{fig5}, wherein it is evident that the plasma oscillation frequency
is increased when the momentum-dependent mass is used because many of the
particles produced are now lighter than the reference mass and hence respond
more quickly to changes in the electric field.

We have sketched a quantum Vlasov equation approach to the study of particle
creation in strong fields.  Using a simple model to mimic a relativistic
heavy ion collision, we solved for the single particle distribution function
and the associated particle currents and electric fields.  Plasma
oscillations are a necessary feature of all such studies.  We illustrated
that the oscillation frequency depends on: the initial energy density reached
in the collision, it increases with increasing energy density; and the
particle-particle collision probability, it decreases as this probability
increases; and that it is sensitive to the necessary momentum-dependence of
dressed-particle mass functions.  The response in each case is easy to
understand intuitively and that is a strength of the Vlasov equation
approach.

\bigskip

\hspace*{-\parindent}{{\large\bf Acknowledgments.}~A.V.P.\ is grateful for
financial support provided by the Deutsche Forschungsgemeinschaft under
project no.\ 436 RUS $17/102/00$.  This work was supported by the US
Department of Energy, Nuclear Physics Division, under contract no.\
W-31-109-ENG-38; the US National Science Foundation under grant no.\
INT-9603385; and benefited from the resources of the National Energy Research
Scientific Computing Center.

\begin{small}

\end{small}
\end{document}